\input harvmac
 
\overfullrule=0pt
\def\Title#1#2{\rightline{#1}\ifx\answ\bigans\nopagenumbers\pageno0\vskip1in
\else\pageno1\vskip.8in\fi \centerline{\titlefont #2}\vskip .5in}

\lref\malda{J. Maldacena, ``The Large $N$ Limit of Superconformal Field
Theories and Supergravity,'' 
{\it Adv. Theor. Math. Phys.} {\bf 2} (1998) 231, hep-th/9711200.}

\lref\bek{J. D. Bekenstein, ``Black Holes and the Second Law,''
{\it Nuovo Cim. Lett.} {\bf 4} (1972) 737.}

\lref\hft{G.~'t Hooft, ``Dimensional Reduction in Quantum Gravity,'' in
{\it Salaamfest 1993,} p. 284,
gr-qc/9310026.}

\lref\susk{ L. Susskind, ``The World as a Hologram,'' 
{\it J. Math. Phys.} {\bf 36} (1995) 6377, 
hep-th/9409089. }

\lref\bfss{T. Banks, W. Fischler, S. Shenker and L. Susskind,
``M-theory as a matrix model: a conjecture,'' {\it Phys. Rev.} {\bf D55}
(1997) 5112, hep-th/9610043.}

\lref\withol{E. Witten, ``Anti-de Sitter space and holography,''
{\it Adv. Theor. Math. Phys.} {\bf 2} (1998) 253,
hep-th/9802150.}

\lref\gkp{S. Gubser, I. Klebanov and A. Polyakov,
``Gauge theory correlators from noncritical string theory,''
{\it Phys. Lett.} {\bf B428} (1998) 105,
hep-th/9802109.}

\lref\witqcd{E. Witten, ``Anti-de Sitter Space, Thermal Phase Transition,
and Confinement In Gauge Theories,'' 
{\it Adv. Theor. Math. Phys.} {\bf 2} (1998) 505, hep-th/9803131.}

\lref\dwk{J. S. Dowker, ``Effective actions on the squashed three-sphere,''
hep-th/9812202.}

\lref\fef{C. L. Fefferman, 
``Monge-Amp\`ere equations, the Bergman kernel, and 
geometry of pseudoconvex domains,''
{\it Ann. of Math.} {\bf 103} (1976) 395.}

\lref\cy{S. Y. Cheng and S.-T. Yau,
``On the Existence of a Complete K\"ahler Metric on Non-Compact Complex
Manifolds and the Regularity of Fefferman's Equation,''
{\it Comm. Pure Appl. Math.} {\bf 33} (1980) 507.}

\lref\bbs{K. Becker, M. Becker and A. Strominger,
``Fivebranes, Membranes and Non-Perturbative String Theory,''
{\it Nucl. Phys.} {\bf B456} (1995) 130,
hep-th/9507158.}

\lref\bsv{M. Bershadsky, V. Sadov and C. Vafa, ``D-branes and
Topological Field Theories, ``
{\it Nucl. Phys.} {\bf B463} (1996) 420, hep-th/9511222.}

\lref\witkleb{I. Klebanov and E. Witten,
``Superconformal Field Theory on Threebranes at a Calabi-Yau Singularity,''
{\it Nucl. Phys.} {\bf B536} (1998) 199, hep-th/9807080; 
S. Gubser, ``Einstein
manifolds and Conformal Field Theory,'' {\it Phys. Rev.} {\bf D59} (1999) 25006,
hep-th/9807164;
M. J. Duff, H. Lu and C. N. Pope, ``$AdS_3 \times S^3$
(Un)twisted and Squashed, and an $O(2,2;Z)$ Multiplet of Dyonic Strings,''
{\it Nucl. Phys.} {\bf B544} (1999) 145,
hep-th/9807173;
B. S. Acharya, J. M. Figueroa-O'Farrill, C. M. Hull and S. Spence,
``Branes at conical singularities and holography,''
{\it Adv. Theor. Math. Phys.} {\bf 2} (1998) 1249, hep-th/9808014;
D. R. Morrison and M. R. Plesser, ``Non-Spherical Horizons, I,''
hep-th/9810201.
}

\lref\romans{L. Romans, ``New Compactifications of Chiral N=2,
d=10 Supergravity,'' {\it Phys. Lett.} {\bf 153B} (1985) 392.}

\lref\crw{L. Castellani, L. J. Romans and N. P. Warner, ``Symmetries of Coset
Spaces and Kaluza-Klein Supergravity, ``
{\it Ann. Phys.} {\bf 157} (1984) 394;
L. Castellani, L. J. Romans and N. P. Warner,
``A Classification of Compactifying Solutions for d=11
Supergravity,'' 
{\it Nucl. Phys.} {\bf B241} (1984) 429.
}

\lref\sugra{B. Biran, A. Casher, F. Englert, M. Rooman and P. Spindel,
``The Fluctuating Seven Sphere in Eleven-Dimensional Supergravity,''
{\it Phys. Lett.} {\bf 134B} (1984) 179.}

\lref\myers{A. Chamblin, R. Emparan, C. Johnson and R. Myers,
``Large $N$ Phases, Gravitational Instantons and the
Nuts and Bolts of AdS Holography,'' 
{\it Phys. Rev.} {\bf D59} (1999) 64010, hep-th/9808177.}

\lref\hawk{S. W. Hawking, C. J. Hunter and D. N. Page,
``Nut Charge, Anti-de Sitter Space and Entropy,'' {\it Phys. Rev.} {\bf D59} (1999)
44033, hep-th/9809035.}

\lref\chg{W. Fischler and L. Susskind, ``Holography and Cosmology,'' 
hep-th/9806039;
N. Kaloper and A. Linde, ``Cosmology vs. Holography,''
hep-th/9904120;
R. Easther and D. A. Lowe, ``Holography, Cosmology and the Second Law of
Thermodynamics,'' hep-th/9902088.}

\lref\hht{S. W. Hawking, C. J. Hunter and M. M. Taylor-Robinson,
``Rotation and the AdS/CFT correspondence,'' {\it Phys. Rev.} {\bf D59} (1999)
064005, hep-th/9811056.}

\lref\bars{I. Bars and Z. Teng, ``The unitary irreducible representations 
of $SU(2,1)$,'' {\it J. Math. Phys.} {\bf 31} (1990) 7.}

\def\ee{\end{equation}}
\def\be{\begin{equation}}
\def\bdz{d\bar{z}}
\def\bz{\bar{z}}
\def\p{\partial}

\def\bw{\bar{w}}
\def\hm{\hat{m}}

\Title{\vbox{\baselineskip12pt
\hbox{hep-th/9905211}
\hbox{HUTP-99/A023}
}}
{\vbox{\centerline{Holography for Coset Spaces}
\smallskip
\centerline{}}}

\centerline{
Ruth Britto-Pacumio,\foot{{\tt britto@boltzmann.harvard.edu}}
Andrew Strominger\foot{{\tt andy@planck.harvard.edu}}
and Anastasia Volovich\foot{{\tt nastya@physics.harvard.edu}}
\footnote{$^{\dagger}$}
{On leave from L. D. Landau Institute for Theoretical Physics,
Kosigina 2, Moscow, Russia}
 }
\bigskip\centerline{Department of Physics}
\centerline{Harvard University}
\centerline{Cambridge, MA 02138}

\vskip .2in
\noindent
\centerline{\bf Abstract}
M/string theory on noncompact, negatively curved, cosets which 
generalize $AdS_{D+1}=SO(D,2)/SO(D,1)$ is considered.  
Holographic descriptions in terms of a conformal 
field theory on the boundary of the spacetime are proposed.
Examples include $SU(2,1)/U(2)$, which is a Euclidean
signature $(4,0)$ space with no supersymmetry, and 
$SO(2,2)/SO(2)$ and $SO(3,2)/SO(3)$, which are Lorentzian  
signature $(4,1)$ and 
$(6,1)$ spaces with eight supersymmetries. 
Qualitatively new features arise due to the 
degenerate nature of the conformal boundary metric. 

\Date{}

\newsec{Introduction}

In Maldacena's $AdS$/CFT duality \malda, M/string theory on 
$AdS$ is equivalent to a field theory on the boundary of 
$AdS$. This is a concrete example of the plausibly much more general 
holographic principle
\refs{\bek \hft - \susk}.
The holographic description of $AdS$ gravity relies on 
very special properties of $AdS$, such as the fact that the 
ratio of the volume and 
surface area approaches a constant at large radius. 
Hence it is far from obvious how the holographic principle can be 
concretely realized in a general setting. Discussions of holography 
in cosmology have appeared in \chg, in flat space 
in \bfss, and in negatively curved spaces other than 
$AdS$ in \refs{\witqcd \myers \hawk-\hht}.  

In this paper we propose a holographic description 
of M/string theory in a family of negatively curved symmetric
spacetimes. $AdS_{D+1}$ can be represented as the coset  
$SO(D,2)/SO(D,1)$.  We consider holography for other 
noncompact cosets{\foot{ Holography for the cases of vacua of the form
$AdS \times X$, where $X$ is a compact coset space,
has been studied in \witkleb.}}, mainly 
$SU(2,1)/U(2)$, which is a 
signature $(4,0)$ space with no supersymmetry, and 
$SO(2,2)/SO(2)$ and $SO(3,2)/SO(3)$, which are 
signature $(4,1)$ and 
$(6,1)$ spaces with eight supersymmetries. There are many other 
similar noncompact cosets. 
These spaces have unusual features such as closed timelike curves but 
nevertheless provide an interesting and challenging 
arena in which to expand our
understanding of holography.

An important new feature is that the conformal boundary metric 
for these cosets has zero eigenvalues. 
This feature also appears in the conformal boundary metric at null infinity 
in Minkowski space, and so may be pertinent in more physically interesting 
spacetimes. Despite the degeneracy of the conformal boundary metric there is 
a nondegenerate conformal boundary measure. We argue that this is enough 
to enable us to define the boundary theory via its correlators.
We find that, as in $AdS$, the bulk isometries become conformal 
isometries of the boundary, and the boundary scalar operators and 
scalar correlation functions transform
accordingly. (Similar results may hold for higher spin, but they are not 
explicitly investigated here.)  
Another generic feature, associated with the 
degeneracy of the boundary metric, is the appearance of an 
infinite-dimensional enlarged conformal symmetry group, 
in some ways analogous to the enlargement of $so(2,2)$ to 
two copies of the Virasoro algebra on the boundary of $AdS_3$. 
 
This paper is organized as follows.
In section 2 we consider M-theory on $SU(2,1)/U(2)\times S^7$. 
The geometry and symmetries of $SU(2,1)/U(2)$ and its conformal boundary
are described. The boundary measure is conformally the standard round
measure 
on $S^3$, while the conformal boundary metric has signature $(+,0,0)$. 
The Einstein-K\"ahler deformations are discussed, following \refs{\fef,\cy}.
The compactification is shown to be free of tachyonic 
instabilities. A prescription is given, 
generalizing \refs{\malda,\withol,\gkp}, 
for defining the correlators of the boundary conformal field theory 
as appropriately rescaled limits of bulk correlators. They are 
seen to be finite despite the degeneracy of the boundary metric. 
Two-point functions are explicitly computed using the  $SU(2,1)$ conformal isometry
group. Section 3 concerns IIB string theory on $SO(2,2)/SO(2)\times S^5$ 
and also briefly M-theory on $SO(3,2)/SO(3)\times S^4$. These are both 
supersymmetric Lorentzian signature spacetimes. In the former case we 
propose that the appropriate boundary theory is two-dimensional. 
In the final section 4
we conjecture a dual description in terms of conformal field theories 
from branes on spacetimes with degenerate metrics. We also describe how 
solitons can spontaneously break $SO(D,2)$ conformal invariance  
down to a smaller subgroup, and suggest that at the duals to such 
configurations may in some cases be interesting Lorentzian cosets.

\newsec{M-theory on $SU(2,1)/U(2)\times S^7$}

In this section we consider M-theory 
compactified
on $SU(2,1)/U(2)\times S^7$ and its holographic representation on 
the boundary of $SU(2,1)/U(2)$.
This is a Euclidean space with no supersymmetry, as can be easily seen 
from the absence of a candidate supergroup. In subsection 2.1 
we describe the bulk geometry of this space as well as the 
degenerate conformal
geometry of the boundary. Relevant results relating the 
metric deformations to boundary data \refs{\fef,\cy} are recalled in 
subsection 2.2.  In 2.3 the mass of 
a scalar field is related to the quadratic Casimir of $SU(2,1)$
and it is shown that there are no tachyonic instabilities. 
In subsection 2.4 a modification of the $AdS$/CFT prescription is 
given for constructing the scalar correlators of the conformal field theory on 
the boundary as limits of bulk correlators.

\subsec{Geometry of  $SU(2,1)/U(2)$ and its conformal boundary}

The coset space $H=SU(2,1)/U(2)$
is topologically the open ball in ${\bf C}^2$
with the Bergman metric
\eqn\mber{
d s^2={dz_1 \bdz_1+dz_2 \bdz_2 \over 1-z_1 \bz_1-z_2 \bz_2 }
+{1 \over (1-z_1 \bz_1-z_2 \bz_2)^2 }
(\bz_1 dz_1 + \bz_2 dz_2)(z_1 \bdz_1+z_2 \bdz_2),}
where $z_1 \bz_1+z_2 \bz_2<1$. This is a K\"ahler metric with K\"ahler 
potential
\eqn\drt{K=-\half \ln (1-z_1 \bz_1-z_2 \bz_2).}
Under the change of coordinates
\eqn\chz{
z_1=r \cos{{\theta \over 2}} e^{i(\varphi+\psi)/2},~~~~~
z_2=r \sin{{\theta \over 2}} e^{-i(\varphi-\psi)/2},}
this metric takes the form
\eqn\sigmet{ds^2={dr^2 \over (1-r^2)^2}+{r^2 \over 4(1-r^2)}{(\sigma_1^2+
\sigma_2^2)}+{r^2 \over 4(1-r^2)^2} \sigma_3^2,}
where the left-invariant one-forms are 
\eqn\sigmas{\eqalign{
\sigma_1&=\cos\psi d\theta+\sin\psi \sin\theta d\varphi,~~\cr
\sigma_2&=-\sin\psi d\theta+\cos\psi \sin\theta d\varphi,~~\cr
\sigma_3&=d\psi+\cos\theta d\varphi.}}
In this metric  $r \in [0,1), \theta \in [0,\pi), \varphi \in [0, 2\pi),$ and $\psi \in
[0,4\pi).$ 
Defining $r=\tanh y$ yields yet another form of the metric,
\eqn\ymetric{
ds^2 = dy^2 + {1 \over 4}\sinh^2y (\sigma_1^2 + \sigma_2^2) + {1 \over 4}\sinh^2
y \cosh^2y \sigma_3^2.}

The geometry \mber\ has an $SU(2,1)$ isometry group 
because the left action on the $SU(2,1)$ group manifold 
remains unbroken in the quotient by the right action of $U(2)$. 
This group is generated by the following eight
Killing vectors. 
\eqn\cch{
H_1=z_1 \partial_{z_1}-\bz_1 \partial_{\bz_1},~~~
~~H_2=z_2 \partial_{z_2}-\bz_2 \partial_{\bz_2},}
\eqn\cclone{
L_1=z_2 \partial_{z_1}-\bz_1 \partial_{\bz_2}, ~~~~~
\bar{L}_1=\bz_2 \partial_{\bz_1}-z_1 \partial_{z_2},
}
\eqn\ccltwo{
L_2=\partial_{z_1}-\bz_1 \bz_2 \partial_{\bz_2}-\bz_1^2 \partial_{\bz_1},
~~~~~\bar{L}_2=\partial_{\bz_1}-z_1 z_2 \partial_{z_2}-z_1^2
\partial_{z_1},}
\eqn\cclthree{{L_3}=\partial_{\bz_2}-z_1 z_2 \partial_{z_1}-z_2^2
\partial_{z_2},~~~~~
\bar{L}_3=\partial_{z_2}-\bz_1 \bz_2 \partial_{\bz_1}-\bz_2^2
\partial_{\bz_2}.
}
The commutation relations between these generators are given in appendix A.
So far the structure of $SU(2,1)/U(2)$ is qualitatively similar 
to Euclidean $AdS_4$, which is the coset $SO(4,1)/SO(4)$. 
However, the structure of the conformal boundary is quite different. 
The conformal boundary metric is determined (up to conformal transformations) 
by rescaling \ymetric\ by a singular 
function of $y$ such that the induced metric 
at the boundary $y=\infty$  is finite. 
Rescaling \ymetric\ by $64 e^{-4y}$ yields the induced metric 
on a surface of constant $y$,
\eqn\indmet{ds^2=4 e^{-2y}(1-e^{-2y})^2(\sigma_1^2 +\sigma_2^2)+(1-e^{-4y})^2
\sigma_3^2 .}
This is a squashed three-sphere. As the boundary is approached,
the squashing becomes more and more severe, until finally 
at the boundary it degenerates to 
\eqn\dgn{ ds^2=\sigma_3^2=(d \psi+\cos{\theta} d\varphi)^2.}
This metric has signature $(+,0,0)$.

Degenerate conformal metrics have appeared in other contexts. 
For example, the boundary of $AdS_4\times S^7$ is $S^3\times S^7$, 
but after conformal rescaling, the metric on the $S^7$ factor 
is degenerate, and one has an effectively three-dimensional metric. 
An analogous interpretation of \dgn\ as a metric on a one-dimensional 
space does not seem possible, since the one-form $\sigma_3$ is not closed. 
Another example is the conformal metric at null infinity of Minkowski
space, which has signature $(0,+,+)$. This last  example suggests 
that the problem of degenerate boundary metrics may be relevant 
for flat space holography. 

Since the metric \dgn\ is degenerate, the associated measure on the
boundary vanishes. It is nevertheless possible to define a conformal
measure on the boundary. Rescaling \ymetric\ by 
$2^{14/3}e^{-8y/3}$, one finds the finite induced volume form at 
the boundary
\eqn\msr{\epsilon_3=\sigma_1\wedge \sigma_2\wedge \sigma_3 ,}
and associated measure
\eqn\msrt{d^3\Omega=\sin \theta d\theta d\psi d\phi.}
Global scale transformations in the boundary theory are induced by shifts 
of $y$. 
Since different powers of $e^y$ are required to make the induced measure
and metric finite, their scaling dimensions will not be related by the
usual factor of $2/3$ (in three dimensions). 
Rather the scale transformations are
\eqn\fgt{\eqalign{ds_3^2&\to \Omega^2(\hat x) ds^2_3, \cr \epsilon_3&\to
\Omega^{2}(\hat x) 
\epsilon_3, \cr }}
where $\hat x$ is a coordinate on the $S^3$ boundary.

Despite the degeneracy of the metric, the conformal Killing equation 
\eqn\keq{{\cal L}_\xi g_{ab}=f(\hat x)g_{ab},}
 which does not involve the 
inverse metric, is well-defined. 
Conformal 
Killing vectors on the boundary with an $su(2,1)$ Lie bracket algebra 
are obtained by  restrictions of \cch--\cclthree,
namely 
\eqn\h{
h_1 =-i(\p_{\varphi}+\p_{\psi}), ~~~~  h_2= i(\p_{\varphi}-\p_{\psi}),}
\eqn\lone{
l_1=-e^{-i\varphi}(\p_{\theta}+{i \over \sin\theta}
(\p_{\psi}-\cos{\theta} \p_{\varphi})),
~~~\bar{l}_1=-e^{i\varphi}(\p_{\theta}-{i \over \sin\theta}
(\p_{\psi}-\cos{\theta} \p_{\varphi})),}
\eqn\ltwo{
l_2=-e^{-i(\varphi+\psi)/2}(\sin{{\theta \over 2}}\p_{\theta}+
{i \over 2 \cos{{\theta \over 2}}}
(\p_{\varphi}+(1+2\cos^2{{\theta \over 2}})
\p_{\psi})),}
\eqn\bltwo{
\bar{l}_2=-e^{i(\varphi+\psi)/2}(\sin{\theta \over 2}\p_{\theta}-
{i \over 2 \cos{\theta \over 2}}
(\p_{\varphi}+(1+2\cos^2{\theta \over 2})
\p_{\psi})),}
\eqn\lthree{
{l}_3=e^{-i(\varphi-\psi)/2}(\cos{{\theta \over 2}}\p_{\theta}-
{i \over 2 \sin{{\theta \over 2}}}
(\p_{\varphi}-(1+2\sin^2{{\theta \over 2}})
\p_{\psi})).}
\eqn\blthree{
\bar{l}_3=e^{i(\varphi-\psi)/2}(\cos{{\theta \over 2}}\p_{\theta}+
{i \over 2 \sin{{\theta \over 2}}}
(\p_{\varphi}-(1+2\sin^2{{\theta \over 2}})
\p_{\psi}))}
One may check explicitly that the function $f$ in \keq\ is 
\eqn\lokj{f(\hat x)=\nabla_m\xi^m.}
$f$ vanishes for  \h--\lone\ which are the $SU(2)\times U(1)$
isometries of the boundary.
We note that despite the degeneracy of the metric the covariant 
divergence is still well-defined. When the metric is nondegenerate, 
one can easily show that 
the coefficient on the right hand side of \keq\ is always 
$2 \over D$ (in $D$ dimensions) with $f$ defined in \lokj,
simply by contraction with the inverse metric.  
However, the metric \dgn\ does not have an inverse so no such demonstration
is possible, and we remarkably find the same function with a different 
coefficient.  
These conformal Killing vectors also preserve the measure
\eqn\mgh{{\cal L}_\xi \epsilon_3=f(\hat x)\epsilon_3.}
Here we encounter the standard conformal transformation law for a
nondegenerate measure.

Due to the degeneracy of the metric, there are infinitely many conformal
Killing vectors in addition to \h--\blthree, as detailed in appendix B. 
Because they do not arise from the isometries of the bulk, there is no
reason to expect that they annihilate the vacuum or provide simple 
relations among the correlators of the boundary theory. A somewhat 
similar situation occurs in $AdS_3$, for which the bulk isometries 
are $SL(2,R)\times SL(2,R)$, but the conformal Killing vectors of the 
boundary theory generate two copies of the Virasoro algebra. In that case
the existence of the infinite-dimensional Virasoro algebra of course 
has profound consequences for the boundary theory. We do not know if 
that is also the case for $SU(2,1)/U(2)$.   

\subsec{Deformations}
   
In this section we discuss deformations of the 
metric $SU(2,1)/U(2)$ which preserve the Einstein equations but in general
destroy the isometries. There is a well-developed theory 
of such deformations. A key relevant result \refs{\fef,\cy} 
is that for any strictly pseudoconvex domain $\Omega$ in ${\bf C}^n$ with a smooth boundary 
$\p \Omega$
there is a $\it unique$ complete Einstein-K\"ahler metric. 
The K\"ahler potential is\foot{The factor of $\half$, not present in 
\cy , is inserted to conform to the conventions of this paper.} 
\eqn\khl{K=-\half \ln s,}
where $s$ is a solution of Fefferman's equation (slightly rewritten)
\eqn\feq{ (-s)^{n+1}\det \p_{i}\p_{\bar j}\ln s =-1} 
subject to the Dirichlet boundary condition
\eqn\dbc{s|_{\p \Omega}=0.}
This boundary condition ensures that the boundary points are an infinite 
distance from the interior. 

The case of $SU(2,1)/U(2)$ arises from the domain in ${\bf C}^2$ bounded by
the $S^3$ given by
\eqn\dmn{v\equiv  1 - |z_1|^2 - |z_2|^2 = 0.}
The solution of \feq\ is then simply 
\eqn\rdt{s= v.}
It is easy to check that the resulting K\"ahler metric is indeed 
the Bergman metric \mber. 

A K\"ahler-Einstein deformation of the Bergman metric can then be 
succinctly described by
deforming the equation for the boundary \dmn, for example by a 
polynomial in $(z, \bar z)$.  One can then find $s$ 
near the boundary in a power 
series expansion in $v$ with a $\ln v$ term.

\subsec{Scalar fields and stability}

Consider a scalar field $\phi$ with mass $m$. The wave 
equation is 
\eqn\wefq{\nabla^2\phi-m^2\phi=0}
where $\nabla^2$ is the Laplacian for the metric \sigmet,
\eqn\laplac{\eqalign{
\nabla^2=
(1-r^2)^2~\p_{rr} + {(1-r^2)(3-r^2) \over r}~ \p_{r} +
~~~~~~~~~~~~\cr
{4(1-r^2) \over r^2}
(\p_{\theta \theta} + \cot{\theta} \p_{\theta}
+ \csc^2{\theta}( \p_{\varphi \varphi}-2\cos{\theta} \p_{\varphi \psi} + 
 (1-r^2 \sin^2{\theta}) \p_{\psi \psi}).
}}
The quadratic Casimir for $SU(2,1)$ is
\eqn\cas{
C_{II}= 
-{1 \over 2}\{L_1,\bar{L}_1\}+
{1 \over 2}\{L_2,\bar{L}_2\}+
{1 \over 2}\{L_3,\bar{L}_3\}+T^2+{3 \over 4}Y^2
,}
where we
redefined $Y=-H_1-H_2, ~T=(H_2-H_1)/2.$
Using the vector fields \cch--\cclthree, we
find that the Laplacian is proportional to the Casimir with a 
factor of 4.   Therefore the solutions of the wave equation for a 
scalar field of mass $m$ form a representation of 
$SU(2,1)$ with quadratic Casimir
\eqn\laplcasi{C_{II}={m^2 \over 4}.}

Next we use the $su(2,1)$ algebra to classify the
solutions of this equation. The representations were studied in \bars. 
The rank of $SU(2,1)$ is two and highest-weight representations are labelled
by  $(t,y),$ such that
 \eqn\hws{Y |\psi\rangle=y |\psi\rangle, ~~~~~T |\psi\rangle =t |\psi\rangle.}
Using the commutation relations
~~$[L_1,\bar{L}_1 ]= -2T,$ ~~$[L_2, \bar{L}_2]=-{3 \over 2}Y+T,$~~ 
$[L_3, \bar{L}_3]={3 \over 2}Y-T,$ ~~
and the highest-weight conditions
\eqn\annih{
L_1 |\psi\rangle=
L_2 |\psi\rangle=
L_3 |\psi\rangle=0,}
we obtain from \cas\ the equation
\eqn\expF{
C_{II}|\psi\rangle = t^2 + 2t +{3 \over 4}y^2 |\psi\rangle.}
or in terms of integers $(p,q)$ such that
$t={1 \over 2}(p+q)$ and $y={1 \over 3}(p-q)$,
\eqn\mpq{
m^2 = {4 \over 3}(p^2 + q^2+pq)+4(p+q).}
For the scalar field $\phi$, the highest-weight conditions 
imply $y=0,$ or $p=q$
and we obtain the relation between the mass $m$ of the scalar field 
and the highest weight $p$ of the form 
\eqn\scalarm{m^2=4p(p+2).}

The functional integral in the quantum theory includes all 
normalizable modes of $\phi$,  even if they do not solve the wave
equation. 
These can be characterized as eigenmodes 
$\phi_k$ of the Laplacian with eigenvalues $\lambda_k$
that obey 
\eqn\wth{-\nabla^2 \phi_k+m^2 \phi_k =\lambda_k\phi_k.}
If there is a negative eigenvalue 
$\lambda_k$ with normalizable eigenmode $\phi_k$, fluctuations of 
$\phi_k$ are unstable. We wish to show that no such instabilities arise 
for M-theory on $SU(2,1)/U(2)\times S^7$. Supersymmetry cannot be 
invoked since there are no appropriate 
covariantly constant spinors in this geometry. 

At large $y$, the angular part of the Laplacian is 
exponentially suppressed, and $\phi_k$ 
obeys
\eqn\rty{{1 \over \sqrt{g}}\p_y \sqrt{g}\p_y \phi_k 
=e^{-4y}\p_y e^{4y}\p_y \phi_k= (m^2-\lambda_k)\phi_k,}
using $\sqrt{g}\to e^{4y}$.
This implies that the leading asymptotic behavior of $\phi_k$ is 
\eqn\ash{\phi_k \to e^{(-2 +\sqrt{4+m^2 -\lambda_k} )y}.}
On the other hand, if $\phi_k$ is normalizable
we need 
\eqn\rto{\phi_k \le e^{-2y}} 
at infinity. It is possible to 
satisfy \rto\ with negative $\lambda_k$ only if 
$m^2<-4$.
\foot{We assumed here that, as is generically expected,
the large $y$ behavior is governed by the dominant exponent \ash;
in principle the coefficient of this term could vanish.}
 The spectrum of eleven-dimensional supergravity on $S^7$
has been studied in \sugra\ and has been shown to contain three families
of scalars with masses $m^2={1\over 4}((k-3)^2-9);
~{1\over 4}((k+8)^2-9); ~{1\over 4}((k+3)^2-9);$
and two families of pseudoscalars with masses
$m^2={1 \over 4}(k^2-9); ~{1\over 4}((k+6)^2-9);$
where $k=1,2,..$ etc.  
(Here we have shifted the mass and performed the overall
normalization so that  conventions of \sugra\ 
agree with those in \withol).
The most negative mass$^2$ is $m^2=-{9 \over 4}$, which is 
insufficient to produce an instability.

The theory also contains vector fields with linearized equations 
of motion
\eqn\lom{d*dA=0.}
Consider an ansatz
\eqn\lio{A=a(y)\sigma_1.}
Then \lom\ becomes 
\eqn\klp{(\p_y^2 a+2\p_y a)e^{2y}dy\sigma_2\sigma_3+{\cal O}(e^{-2y})dy\sigma_2\sigma_3=0.}
This implies that the leading behavior of $A$ is a constant, and 
that there is no normalizable zero mode of the form \lio. Similar
conclusions apply to $A\sim \sigma_2$, while for $A=a\sigma_3$ 
one finds $a \sim e^y$, which is also non-normalizable. 
Hence there are no normalizable zero modes of this form. 
Allowing for angular dependence of $a$, a negative eigenvalue 
for \lom, or considering massive vectors in the theory 
(which all have $m^2>0$) only makes it more difficult to get a 
normalizable eigenmode.  In conclusion, the vector fields on 
$SU(2,1)/U(2)$ also do not induce an instability. 
We have also checked that normalizable graviton zero modes
do not exist, in harmony
with
the uniqueness theorem \cy\ discussed in section 2.2. 
We conclude that $SU(2,1)/U(2)\times S^7$ 
is a stable solution of M-theory.

\subsec{The Boundary Theory}

In accord with the holographic principle, we wish to represent the bulk  
M-theory on  $SU(2,1)/U(2)\times S^7$ as a conformal 
field theory on the conformal boundary of $SU(2,1)/U(2)\times S^7$.
In this subsection we describe (for scalars) 
how the operators and correlation 
functions of this boundary theory can be defined as limits of 
various bulk quantities. This procedure is a modification of that used to 
define the boundary theory for $AdS_4\times S^7$.  The 
resulting correlators are well-behaved and transform appropriately under
the  boundary conformal group $SU(2,1)$, despite the degeneracy of the 
boundary metric.  In section 4
we discuss possible dual representations in terms of branes. 

Let us consider the conformal field theory on the 
boundary of $SU(2,1)/U(2)\times S^7$
with the degenerate metric 
\eqn\metrd{ds^2=\sigma_3^2=(d \psi+\cos{\theta} d\varphi)^2,}
and measure
\eqn\msr{d^3\Omega =\sin \theta d\theta d\psi d\phi.}
A conformal transformation is a diffeomorphism together with 
a Weyl transformation. A field of conformal dimension $\Delta$
transforms as
\eqn\varphitr{
\delta_\xi {\cal O} = ({\cal L}_\xi+{\Delta \over 3}\nabla_m \xi^m) 
{\cal O},}
where ${\cal L}_\xi$ is the usual Lie derivative, equal to $\xi^m \p_m$
acting on scalars. The metric and measure both have $\Delta=-3$. 

For simplicity let us restrict our attention to scalar operators 
$\cal O$ in the boundary. Let $\delta_i$, with $i=1,2,\ldots,8,$ 
denote the eight $SU(2,1)$ conformal transformations generated by
the vectors \h--\blthree\ on $S^3$.
The quadratic Casimir associated to such an operator 
follows from squaring \varphitr\ as  
\eqn\csm{C_{II}{\cal O}=g^{ij}\delta_i\delta_j {\cal O}={4\over 9}
\Delta(\Delta-3){\cal O},}
where $g_{ij}$ is the flat signature $(4,4)$ metric for the $su(2,1)$ 
Lie algebra appearing in \cas. 
Comparing \csm\ to \laplcasi\ we see that for every scalar field of mass $m$ there
is a boundary operator with weight $\Delta$ obeying
\eqn\msg{m^2={16\over 9}
\Delta(\Delta-3).}

The two-point function of the scalar fields
$\langle{\cal O}_{\Delta_1}(z){\cal O}_{\Delta_2}(w)\rangle$
is fixed by the requirement of invariance under conformal
transformations. The requirement of invariance under the
isometries generated by \h--\lone\ 
leads to the following equations:
\eqn\eqh{\eqalign{
[h_1^{(z)}+h_1^{(w)}] \langle{\cal O}_{\Delta_1}(z)
{\cal O}_{\Delta_2}(w)\rangle&=0, ~~~\cr
[h_2^{(z)}+h_2^{(w)}] \langle{\cal O}_{\Delta_1}(z){\cal O}_{\Delta_2}(w)
\rangle&=0,\cr
[l_1^{(z)}+l_1^{(w)}] \langle{\cal O}_{\Delta_1}(z)
{\cal O}_{\Delta_2}(w)\rangle&=0, ~~~\cr
[\bar{l_1}^{(z)}+\bar{l_1}^{(w)}]
\langle{\cal O}_{\Delta_1}(z){\cal O}_{\Delta_2}(w)\rangle&=0,}}
where the superscripts $(z)$ and $(w)$
on the generators denote the coordinates
on $S^3.$ In order to fully exploit the symmetries, 
the  $S^3$ coordinate $z$ is traded for an 
$SU(2)$ group element $g$ defined by 
\eqn\g{
g_z=
\pmatrix{z_1 & \bz_2 \cr -z_2 & \bz_1}
=\pmatrix{
\cos{{\theta \over 2}} e^{i(\varphi+\psi)/2}
& \sin{{\theta \over 2}} e^{i(\varphi-\psi)/2} \cr
  -\sin{{{\theta \over 2}}} e^{-i(\varphi-\psi)/2} & \cos{{\theta \over 2}}
  e^{-i(\varphi+\psi)/2}}.}
Invariance under the $SU(2)\times U(1)$ isometries generated by 
\eqh\ then requires that the correlators are invariant under a left
$SU(2)$ action and a right $U(1)$ action on $g$.  
This requires that the correlator depends only on 
two real functions or one complex function 
$U$ 
\eqn\wht{\langle{\cal O}_{\Delta_1}(z){\cal O}_{\Delta_2}(w)\rangle=f(U,
\bar{U}),}
where 
\eqn\lkj{\eqalign{U&={1 \over 2}\Tr[(1+\sigma_3)g_z^\dagger g_w ]\cr &=
\bz_1 w_1+\bz_2 w_2\cr &=
\cos {\theta_z \over 2}\cos {\theta_w \over 2}
\exp{{i(\varphi_w+\psi_w-\varphi_z-\psi_z) \over 2}}\cr&~~~~~~+
\sin {\theta_z \over 2}\sin {\theta_w \over 2}
\exp{{-i(\varphi_w-\psi_w-\varphi_z+\psi_z) \over 2}},} }
and $\bar{U}$ is the conjugate.
The requirement for  the two-point function   
to be covariant 
under the transformation generated by $l_2$ 
is
\eqn\eqltwo{
[l_2^{(z)}+l_2^{(w)}] \langle{\cal O}_{\Delta_1}(z) 
{\cal O}_{\Delta_2}(w)\rangle=
-{1 \over 3}[\Delta_1 \nabla \cdot l_2^{(z)}+\Delta_2 \nabla \cdot l_2^{(w)}]
\langle{\cal O}_{\Delta_1}(z) {\cal O}_{\Delta_2}(w)\rangle,}
which can be rewritten in the form
\eqn\eq{
\bz_1 (-{2 \over 3}\Delta f +(1-U) \p_U f)+ 
\bw_1 (-{2 \over 3}\Delta f+(1-\bar{U})
\p_{\bar{U}}f)=0,}
where $\Delta=\Delta_1=\Delta_2.$  The condition that $\Delta_1=\Delta_2$
follows from comparing equation \eqltwo\ with its conjugate.
Note that $\nabla \cdot l_2 \equiv {1 \over \sqrt{g}}\p_i (\sqrt{g}~ l_2^i)=
2 \bar{z}_1$,
where $\sqrt{g} \sim \sin \theta.$

The other three equations have
$(z_1,w_1),$  $(\bz_2,\bw_2)$ and $(z_2, w_2)$ consecutively,
in place of $(\bz_1,\bw_1)$.
The function
\eqn\f{
f(U,\bar{U})=|1-U|^{ -{4\Delta \over 3} }}
satisfies
\eq.  Since
moreover each term in front of
$\bz_1$ and $\bw_1$ vanishes separately,
this function satisfies evidently all the other equations.
Thus we have found that the two-point function of two scalar fields of 
dimension $\Delta$ is given by
\eqn\answer{
\langle{\cal O}_{\Delta}(z) {\cal O}_{\Delta}(w)\rangle =
{const \over |1-U|^{4\Delta \over 3}}.}

In the preceding we saw that conformal invariance determines the 
two-point functions of the boundary
operators. Higher-point functions will not be fully determined by conformal
invariance. 
The recipe for calculating a general correlation function within the $AdS$/CFT
correspondence, 
as formulated in \refs{\withol,\gkp},
is the following: first, compute the supergravity partition function in
terms of
the boundary values of the fields; then, identify the operators in the
boundary conformal field theory whose sources are the given boundary
values; finally, interpret the supergravity partition function as a
generating functional of those operators.  
This prescription associates to each field $\phi$ in the supergravity
action a corresponding operator ${\cal O}$ by the relation
\eqn\adscft{\langle e^{\int_{\partial} \phi_0 {\cal O}}\rangle=e^{-I(\phi)}.}
Here $I(\phi)$
is the classical action evaluated on the 
solutions of the supergravity equation of motion subject to
some boundary condition, the integral is over the boundary, 
and the left hand side is interpreted
as a partition function of the connected Green functions
for the operators ${\cal O}.$  Because of the boundary degeneracy, it is not manifestly obvious 
that this prescription can be adapted to $SU(2,1)/U(2)$. 
In this subsection we see that the divergences cancel and 
the prescription can indeed be adapted.

Let us consider a scalar field of mass $m$ in the bulk of
$SU(2,1)/U(2).$ We will not keep track of finite normalization constants in
the rest of this section. 
In order to compute the correlation
function of the operators ${\cal O},$
we first have to calculate the 
 action 
\eqn\clas{
I(\phi)=  \int dr d\theta d\psi d\varphi  \sqrt{g} 
((\nabla \phi)^2 +m^2 \phi^2)}
for a solution of a classical equation of motion
\eqn\ur{\nabla^2 \phi=m^2 \phi,}
subject to the boundary condition
\eqn\bound{\lim_{r \to 1}\phi(r,\theta,\psi,\varphi)=
(1-r^2)^{2-{2 \Delta \over 3}} \phi_{0}(\theta,\psi,\varphi).}
We use the metric \sigmet\ in the bulk of $SU(2,1)/U(2)$
and the relation between the mass $m$ of the
scalar field and the dimension $\Delta$ of the boundary operator,
$m^2={16 \over 9}\Delta(\Delta-3).$
The solution of \ur\ is given by
\eqn\sol{\phi(r,\theta,\psi,\varphi)=
\int K(r,\theta,\psi,\varphi;\theta',\psi',\varphi') 
\phi_0(\theta',\psi',\varphi') \sin{\theta'}d\theta'd\psi'd\varphi',}
where the bulk-to-boundary propagator
is
\eqn\K{K(r,\theta,\psi,\varphi;\theta',\psi',\varphi') =
{(1-r^2)^{{2 \Delta \over 3}} \over |1-rU|^{{4 \Delta \over 3}} }}
with $U=\cos {\theta \over 2}\cos {\theta' \over 2}
\exp{{i(\varphi'+\psi'-\varphi-\psi) \over 2}}+
\sin {\theta \over 2}\sin {\theta' \over 2}
\exp{{-i(\varphi'-\psi'-\varphi+\psi) \over 2}},$ as above.
Note that 
\eqn\bonk{
\lim_{r \to 1}K(r,\theta,\psi,\varphi;\theta',\psi',\varphi') = {(1-r^2)^{2-{2 \Delta \over 3}}\over \sin{\theta'}} \delta({\theta'-\theta})
\delta({\psi'-\psi})\delta({\varphi'-\varphi}).}
Upon integrating by parts, we find that only the boundary term
contributes to the action \clas:
\eqn\act{I(\phi) = \lim_{r \to 1}
\int d^3 \Omega {1 \over (1-r^2)} \phi \partial_r \phi.}
\eqn\otvet{I(\phi) = -\int d^3 \Omega d^3 \Omega' 
{\phi_0(\theta,\varphi,\psi) 
\phi_0(\theta',\varphi',\psi') \over
|1-U|^{{4\Delta \over 3}}}.}
We see that the boundary action is indeed a finite function of 
$\phi_0$, despite the degeneracy of the boundary metric, 
and that it correctly reproduces the two-point 
function of $\cal O$ as determined by conformal invariance in the 
preceding subsection. In principle this boundary action can also 
be used to determine the higher-point correlation functions of 
$\cal O$ and might also be extended to fields of higher spin.

\newsec{Supersymmetric Lorentzian Cosets}

In this section we consider compactifications of IIB on
$SO(2,2)/SO(2) \times S^5$ and 
M-theory on $SO(3,2)/SO(3) \times S^4,$
where  
$SO(2,2)/SO(2)\equiv W_{4,2}$ and $SO(3,2)/SO(3)\equiv W_{5,2}$
are the noncompact cousins of the Stiefel
manifolds $SO(4)/SO(2)\equiv V_{4,2}$ and $SO(5)/SO(3)\equiv V_{5,2}$. 
Each of these spaces is defined with the divisor subgroup embedded canonically in the larger group.
These examples differ from that of the previous section in that 
they have Lorentzian signature and are supersymmetric. 
The unbroken supersymmetries are described in subsection 3.1. 
The geometry of $W_{4,2}$ and its conformal boundary are detailed in 
subsection 3.2. In 3.3 scalar fields in $W_{4,2}$ 
are described.

\subsec{Supersymmetry }

In this subsection it is shown that the spaces $W_{5,2}\times S^7$ 
and $W_{4,2}\times S^5$ preserve the same amount of supersymmetry
as $AdS_4\times V_{5,2} $ and $AdS_5 \times V_{4,2} $, respectively,
namely eight supersymmetries in all cases.

On a space with nonvanishing cosmological constant, 
unbroken supersymmetries are constructed from solutions of the 
Killing spinor equation
\eqn\kileq{D_m \eta=0,} 
where
$D_m=\nabla_m-i\Gamma_m.$ The integrability condition 
for this equation is that the operator
$[D_m, D_n]={1\over{4}}{C_{mn}}^{ab}\Gamma_{ab}$
has zero modes, where ${C_{mn}}^{ab}$
is the Weyl tensor.  Hence we are interested in the holonomy of the 
Weyl tensor. 

We first recall that the Weyl holonomy of $V_{5,2}$ is $SU(3).$
Following the conventions and methodology of \crw,
define 
${(T^{AB})}_{CD} = \delta^{A}_{C} 
\delta^{B}_{D}-\delta^{A}_{D}\delta^{B}_{C}$
to be the generators of $SO(5),$
where $A, B, C, D$ range from 1 to 5.  
To make the canonical embedding of $SO(3)$ 
($\vec{5} \to \vec{3}+\vec{1}+\vec{1}$)
manifest, rewrite the generators as 
\eqn\x{X^i={1 \over 2}\epsilon^{ijk}T^{jk},
 X^m = T^{4m}, X^{\hm} = T^{5m}, 
X^0 =T^{45},}
where $X^i$ generate the $SO(3)$ subgroup, and  
the indices $i, m, \hm$ range from $1$ to $3.$  
The values of the nonvanishing structure constants
of $SO(5)$ (defined by $[T_A, T_B] = C_{AB}^C T_C$)
are then
\eqn\nchsign{
C_{ij}^k=\epsilon_{ijk}, ~~C_{im}^n=C_{i\hat{m}}^{\hat{n}}=
-\epsilon_{imn}, ~~C^{\hat{n}}_{m0}=-C^{n}_{\hat{m}0}=\delta_m^n,}
\eqn\chsign{C^{i}_{mn}=C^{i}_{\hat{m} \hat{n}}=-\epsilon_{imn}, 
~~~C^0_{m \hat{n}}=-\delta^m_n.}

The metric on $G/H$ inherited from the group-invariant metric on $G$ 
is not in general an Einstein metric, but can sometimes 
be transformed into an 
Einstein metric without losing any isometries by 
appropriately rescaling the vielbein components. 
Consider the matrices of structure constants
$(C_D)_b^{~a}$, as $D$ runs over the indices in the normalizer of
the subgroup, and $a$
and $b$ run over flat coset indices.  It was shown in \crw\ that 
if these matrices are
block diagonal in the spaces spanned by the vielbein components
$e^{a_1},e^{a_2},\ldots,$ then an arbitrary rescaling of the
vielbein, $e^{a_i} \to r(a_i)e^{a_i},$
preserves the original isometries.  One can try to find a rescaling to obtain
an Einstein metric on the coset space.  For $V_{5,2},$ rescale with
$r(m)=r(\hat{m})=4$ and $r(0)=\sqrt{32 \over 3}.$

The Riemann tensor for the rescaled coset can be calculated using the
Maurer-Cartan equations and the Jacobi identities for the products of
structure constants.  In terms of the structure constants and squashing
parameters, the Riemann tensor is
\eqn\riem{\eqalign{ R^a{}_{bde} = {1 \over 4} C_{bc}^a C_{de}^c 
      \left( {a\ b \atop c} \right) {r(d)r(e) \over r(c)}+ 
 {1 \over 2} C_{bi}^a C_{de}^i r(d) r(e) +
\cr
   {1 \over 8} C_{cd}^a C_{be}^c \left( {a\ c \atop d} \right)
    \left( {b\ c \atop e} \right) - 
   {1 \over 8} C_{ce}^a C_{bd}^c \left( {a\ c \atop e} \right)
 \left( {b\ c \atop d} \right),}}
with
\eqn\abc{
  \left( {a\ b \atop c} \right) \equiv
   {r(a) r(c) \over r(b)} + {r(b) r(c) \over r(a)} -
    {r(a) r(b) \over r(c)},}
where $a,b,\ldots$ are the flat $G/H$ coset indices, namely 
$m,\hat{m},0,$ and $i$ is an
$H$ index. 
The flat metric is defined as $\gamma_{ab} = -C^C_{aD} C^D_{bC}$.
The nonvanishing components of the Weyl tensor 
for $V_{5,2}$
read
\eqn\W{
C^{mn}_{~~~pq}=C^{\hat{m}\hat{n}}_{~~~\hat{p} \hat{q}}=
C^{mn}_{~~~\hat{p} \hat{q}}=
5( \delta^m_p \delta^n_q-\delta^m_q \delta^n_p), ~~~
C^{m \hat{n}}_{~~~p \hat{q}}=2 \delta^m_n \delta^p_q-3\delta^m_q \delta^n_p
-3\delta^m_p \delta^n_q.}

The holonomy of $V_{5,2}$ is $SU(3)$ if there exists a two-form 
$J$ such that
\eqn\J{
C_{AB~~N}^{~~~~M} J^N_{~~P}=C_{AB~~P}^{~~~~N} J^M_{~~N}, 
~~C_{AB~~N}^{~~~~M} J_M^{~~N}=0.
}
The form with the components
\eqn\cj{
J_m^{~n}=J_{\hat{m}}^{~\hat{n}}=0,~~~J_m^{~\hat{n}}=-J^{~m}_{\hat{n}}=
\delta^m_n}
satisfies equations \J\ with the Weyl tensor of \W.
Thus, the holonomy of $V_{5,2}$ is $SU(3).$
The spinor $\vec{8}$ of $spin(7)$ decomposes as 
$\vec{8}=\vec{1}+\vec{1}+\vec{3}+\vec{3}^*$
under $SU(3).$ The two singlets account for two covariantly 
constant spinors on $V_{5,2}$. The full symmetry group for 
M-theory on $AdS_4 \times V_{5,2}$ is $OSp(2,2|2) \times SO(5),$
which has 8 supercharges.

Similar arguments are valid for $V_{4,2},$
obtained by the canonical embedding 
of $SO(2)$ in $SO(4):$
$\vec{4} \to \vec{2}+\vec{1}+\vec{1}.$
Upon calculating the Weyl tensor, one finds that the holonomy of
$V_{4,2}$ is $SU(2)$. This eliminates half the supersymmetries,
but, due to chirality constraints, these can be used 
to construct only 8 supercharges for IIB string 
theory on $AdS_5\times V_{4,2}$ 
\romans . The full symmetry group
is $SU(2,2|1) \times SO(4)$.

Now consider $W_{4,2}$ and $W_{5,2}.$
To obtain the generators of $SO(n-2,2)$ from $SO(n),$
simply multiply the generators $X^m$ and $X^{\hat{m}}$ by $i$,
so that only the structure constants in \chsign\
will change sign while those in \nchsign\ remain the same. Note also that only the $\gamma_{00}$
component of the flat metric changes sign.  From \riem, we find that 
$R^{ab}_{~~cd}(W_{n,2})=-R^{ab}_{~~cd}(V_{n,2}).$
Therefore, $C^{ab}_{~~cd}(W_{n,2})=-C^{ab}_{~~cd}(V_{n,2})$. Since 
the ``0'' components of the Weyl tensor all vanish according to \W, 
the flat metric on the algebra generated by  the Weyl tensors 
of $W_{n,2}$ and $V_{n,2}$ 
are the same. Hence the holonomies of $W_{4,2}$ and $W_{5,2}$
are $SU(2)$ and $SU(3)$ respectively.
The full symmetry groups
of compactifications IIB$|_{{W_{4,2}\times S^5}}$ and
M$|_{W_{5,2}\times S^4}$ 
are
$SU(4|1) \times SO(2,2)$ and $OSp(4|2)\times SO(3,2)$
respectively, both of which have 8 supercharges.

\subsec{Geometry of $W_{4,2}\equiv SO(2,2)/SO(2)$ and its boundary}

The coset space obtained by quotienting $SO(2,2)$  by the $SO(2)$
subgroup is a symmetric Einstein space with negative cosmological
constant and signature $(4,1)$. 
Topologically $W_{4,2}$ is $S^1 \times {\bf R}^4,$
so that $\pi_1(W_{4,2})={\bf Z}.$

The Riemannian metric on 
$W_{4,2}$ can be obtained by an analytic continuation of the 
$V_{4,2}$ metric and takes the form
\eqn\metWf{
ds^2=
-{1 \over 9}(d \psi + \cosh{y_1} d\varphi_1 + \cosh{y_2} d\varphi_2)^2
+{1 \over 6}(d y_1^2 +\sinh^2{y_1} d \varphi_1^2+
d y_2^2 +\sinh^2{y_2} d \varphi_2^2),}
where $y_i \in [0, \infty),~$
$\varphi_i \in [0, 2\pi),~$  and $\psi \in [0, 4 \pi ).$
The coordinate $\psi$ parametrizes the $U(1)$ fiber of $W_{4,2}$
viewed as a $U(1)$ bundle over $AdS_2 \times AdS_2.$
It is accordingly convenient to view the geometry as a Kaluza-Klein 
compactification to the four dimensional space
\eqn\metgf{
d\hat s^2={1 \over 6}(d y_1^2 +\sinh^2{y_1} d \varphi_1^2+
d y_2^2 +\sinh^2{y_2} d \varphi_2^2),} 
with the $U(1)$ gauge field strength
\eqn\gul{F=\epsilon_1+\epsilon_2 ,}
where the $\epsilon_i$ are proportional to the volume elements of the two
$AdS_2$ 
factors in \metgf. The
isometries of this space are
generated by the  
six Killing vectors
\eqn\algW{\eqalign{L_0^i=i\partial_{\varphi_i},~~~~~~~~~~~
~~~~~~~~~ 
\cr
L^{i}_{-1}= i e^{-i\varphi_i}(\coth{y_i}\partial_{\varphi_i}+i\partial_{y_i}),
\cr
L^{i}_{1}= i e^{i\varphi_i}(\coth{y_i}\partial_{\varphi_i}- 
i\partial_{y_i}),
}}
which, together with $J=i \partial_{\psi},$ generate an $so(2,2) \times
so(2)$ algebra 
($so(2,2) \cong sl(2,R) \times sl(2,R)$):
\eqn\alg{
[L_0^i, L_{\pm 1}^j]=\mp \delta^{ij} L_{\pm 1}^i,~~
[L_1^i, L_{-1}^j]=2 \delta^{ij}L_0^i, ~~
[L_0^i, J]=[L_1^i, J]=[L_{-1}^i, J]=0,}
where $i=1,2.$

The boundary of $W_{4,2}$ might be defined by
\eqn\b{ \sinh^2{y_1}+\sinh^2 {y_2}=\Lambda^2 \to \infty,}
which can be written in terms of a new coordinate 
$\chi \in [0, {\pi \over 2})$ as
\eqn\hi{ \sinh{y_1}=\Lambda \cos{\chi}, ~~~~
\sinh{y_2}=\Lambda \sin{\chi}.}
The conformal boundary metric is 
\eqn\bounccc{
ds^2=\cos^2{\chi} d\varphi_1^2+
\sin^2{\chi} d\varphi_2^2.
}

We have not succeeded in making sense of the notion of 
a theory on the boundary \bounccc.\foot{Similar issues 
arise in other examples such as IIB string theory 
on $AdS_2\times AdS_3 \times S^5$.}   
In most locations it is  a degenerate signature $(+,-,0)$ 
metric, but at $\chi=0,{\pi \over 2}$, it degenerates further to 
signature $(+,0,0)$. 

An alternate procedure that yields a smoother result 
is to suppress the $\chi$ coordinate. A motivation for this is 
that distances in the $\chi$ direction are all zero, 
together with those along with the $U(1)$ fiber 
and the $S^5$, in the conformal boundary metric. 
The variable $\chi$ is eliminated in the two-dimensional (rather than three-dimensional) ``boundary'' defined by 
\eqn\btr{ \sinh^2{y_i}=\Lambda_i^2 \to \infty,}
which is simply $T^2$ with the conformal boundary metric 
\eqn\ftp{ ds^2=d\varphi_1^2+d\varphi_2^2 .}
The $so(2,2)$ algebra on the boundary is 
generated by the vector fields
\eqn\confccc{
l_0^i=i \partial_{\varphi_i},~~~ 
l_{-1}^i=ie^{-i \varphi_i}\partial_{\varphi_i},~~~
l_{1}^i=ie^{i \varphi_i}\partial_{\varphi_i}
.}
This algebra can be extended to two copies of the Virasoro algebra with
the generators
\eqn\vir{l_{n}^i=ie^{i n \varphi_i}\partial_{\varphi_i}.}
Hence the boundary theory may be related to a two-dimensional 
conformal field theory.

Note that the vector fields \confccc\ are not conformal Killing vectors 
of the full boundary. Rather they are each conformal Killing vectors of 
one of the two $S^1$ boundary components. This is related to the
appearance
of the two "scale" parameters $\Lambda_i$ defining the boundary in 
\btr.

A novel feature of this spacetime is the existence of closed timelike
curves. Examples are the curves 
$\chi={\pi \over 4}, ~~\varphi_1=\varphi_2$, 
constant $\psi$ and large $y$. Unlike in $AdS_4$ these cannot be eliminated 
by going to the covering space.

\subsec{Scalar Fields in $W_{4,2}$}
In this subsection we derive the relation 
between the mass $m$ of the scalar field in the bulk and
the highest weights $j,~ h_1,~ h_2$
of the $so(2,2) \times so(2)$ 
algebra.

The scalar field $\phi$
in the bulk of $SO(2,2)/SO(2)$
is described by the wave equation
\eqn\waveqs{\nabla^2 \phi=m^2 \phi,}
where $\nabla^2$ is the Laplacian for the metric \metWf,  
\eqn\lapl{\eqalign{
\nabla^2=
3 ({2 \over \sinh^2{y_1}}\partial_{\varphi_1 \varphi_1 }+
   {2 \over \sinh^2{y_2}}\partial_{\varphi_2 \varphi_2}-
{4 \coth{y_1}\over  \sinh{y_1}} \partial_{\varphi_1 \psi} - 
 {4 \coth{y_2} \over \sinh{y_2}} \partial_{\varphi_2 \psi}
+3 \partial_{\psi \psi}  -
\cr 
2 \coth^2{y_1} \partial_{\psi \psi}  
- 2 \coth^2{y_2} \partial_{\psi \psi}   + 
2 \coth{y_1} \partial_{y_1} + 2 \partial_{y_1 y_1}  
+2 \coth{y_2} \partial_{y_2} + 2 \partial_{y_2 y_2})  
.}}
The Laplacian can be written in terms of the Casimir of $so(2,2)\times so(2)$
as
\eqn\casss{
\nabla^2 =-6( \sum_{i=1}^{2} 
({1 \over 2} \{ L_1^i, L_{-1}^i\}-{L_0^{i2}})+{1 \over 2}J^2).
}
Highest weight states $|h\rangle$ are characterized by 
\eqn\hw{L_0^i |h\rangle=h_i |h\rangle,
 ~~~~~J |h\rangle=j|h\rangle, ~~~~~L^i_{1}|h\rangle=0.}
Acting with the Casimir operator \casss\ on the highest weight state
$|h\rangle$ leads to the relation
\eqn\massss{m^2=6(h_1(h_1-1)+h_2(h_2-1) -j^2/2).
}
Hence for every scalar field $\phi$ of mass $m$ we expect 
operators $\cal O$ in the boundary theory with corresponding weights.

\newsec{Brane constructions}

The holographic principle suggests that M/string theory on a 
given space can be represented as a field theory on the boundary of the 
space.  In the preceding section, following the logic of the 
$AdS$/CFT correspondence, the boundary correlators for various cosets 
have been described as limits of bulk correlators. In some of the 
$AdS$ cases, dual description of the boundary theory 
for example as a large $N$ gauge theory, are possible. In this 
section such dual descriptions will be considered for the cases at hand. 

The field theory on the boundary of $AdS_4\times S^7$ can be defined as 
the infrared limit of a theory of M2-branes, or the strong-coupling,
infrared limit of the D2-brane gauge theory. This theory lives on 
$S^3$ with the round metric.  One may consider the same limit 
on $S^3$ with the squashed metric\foot{Free field theory 
partition functions on this space are computed in \dwk.}
\eqn\gth{ds^2=\sigma_3^2+{1 \over a}(\sigma_1^2 +\sigma_2^2).}
It is natural to conjecture that in the limit that 
the squashing parameter $a$ is taken to infinity, 
one obtains the dual description of M-theory on $SU(2,1)/U(2)\times S^7$.
(A similar conjecture was advanced in the context of Taub-Nut where a
finitely squashed $S^3$ is encountered \refs{\myers,\hawk}.) The results of 
section 2 can be regarded as evidence that this limit is
well-defined.\foot{The scalar curvature of the metric \gth\ is 
$R=2-{1 \over 2 a}$, and so is negative for the $SU(2,1)/U(2)$ 
(as well as Taub-Nut) boundary metric. This will lead to 
Coulomb-branch instabilities near the origin 
for the gauge theory scalars due to the $R\phi^2$ coupling. 
Hence the flow into the infrared could be quite nontrivial, and 
there may be subtleties concerning the order in which the 
infrared and $a \to 0$ limit are taken.  }
Similar conjectures for $W_{4,2}$ and $W_{5,2}$ involve 
Yang-Mills theory on a degenerate four-geometry  and 
the $(0,2)$ fivebrane conformal field theory on a degenerate six-geometry.
While perhaps plausible, these descriptions do not seem terribly useful in
their present formulation and 
are therefore unsatisfying. 

It would be illuminating to find these or other 
noncompact coset spaces as near-horizon geometries of brane configurations.
The branes may have nontrivial worldvolume geometry and/or internal field
excitations. The spacetime supergravity solution for such brane
configurations is not in general known. However it may be possible 
in some cases with enough symmetry to find
the near horizon geometry without knowing the full spacetime solution. 
One construction that may lead to noncompact coset 
spaces---although perhaps not the ones explicitly discussed in this 
paper---involves the
spontaneous breakdown of conformal invariance. This can occur in the
presence of solitons. 
The generators of the conformal group $SO(D,2)$ 
of $D$-dimensional Minkowski space are 
\eqn\ctr{v^a=\lambda x^a+b^ax^2-2x^ab\cdot x,}
together with the Poincare generators. 
A scalar field $\phi$ for example
transforms as 
\eqn\sfld{\delta \phi=v^a\p_a\phi+{D-2 \over 2D}\p_a v^a \phi.}
A given expectation for $\phi$ breaks the conformal group down to a
subgroup generated by those $v$'s that annihilate $\phi$ in 
\sfld. Unbroken global scale invariance (generated by $v=\lambda x$)
requires 
\eqn\rol{x^a\p_a\phi=-{D-2 \over 2}\phi,}
so $\phi$ must scale in the specified way with $x$. In general this implies
for $D>2$ that $\phi$ will be singular at the origin. 
Now consider the special conformal transformations parametrized by 
the vector $b^a$ in \ctr. If $\phi$ is invariant under some translations so
that for longitudinal transformations $b^a_L\p_a\phi=0$, then 
\rol\ is necessary and sufficient to ensure invariance under the associated 
special conformal transformations. The transverse transformations
with $b_T^a\p_a \phi \neq 0$ are necessarily broken. 

In summary, if the field configuration $\phi$ scales as \rol\ 
and is invariant under $d$-dimensional Poincare transformations, 
it follows that the conformal group $SO(D,2)$ is broken 
down to $SO(d,2)$. An obvious generalization of this 
statement pertains to the brane worldvolume metric as well as
 higher-rank tensor fields. 
Further conditions should be imposed if supersymmetry is to 
be preserved.

In general, there are many noncompact cosets of which only three examples
were discussed in this paper.  One obvious generalization is to 
quotient by both a left and a right action. There are also many 
ways to spontaneously break conformal invariance with solitons or 
nontrivial induced metrics 
on a brane worldvolume.  
It would be interesting to find a plausible candidate for a dual pair.

\bigskip
\centerline{\bf Acknowledgements}\nobreak
\bigskip

We are grateful to M. Duff, S. Gubser, G. Horowitz, R. Kallosh, 
J. Maldacena, J. Polchinski, 
C. Pope,  M. Spradlin, E. Witten 
and S.-T. Yau for useful conversations. It has come to our attention that 
related ideas have been independently pursued by A. Mikhailov
and also by M. Taylor-Robinson. 
This work was supported in part 
by DOE grant DE-FGO2-91ER40654 and an NDSEG graduate fellowship. 

\appendix{A}{$SU(2,1)$ commutation relations}

The relations between the generators
\cch\--\cclthree\ and the standard ones $F_i$ are 
\eqn\fos{\eqalign{
F_1={1 \over 2}(L_1 -\bar{L}_1),~~ 
F_2=-{i \over 2}(L_1 +\bar{L}_1),~~
F_3={1 \over 2}(H_2 -H_1),~~
F_8=-{\sqrt{3} \over 2}(H_1+H_2),
\cr
F_4=-{1 \over 2}(L_2 +\bar{L}_2),~~
F_5={i \over 2}(L_2 -\bar{L}_2),~~
F_6={1 \over 2}(L_3 +\bar{L}_3),~~
F_7=-{i \over 2}(L_3 -\bar{L}_3).}}
The standard generators $F_i$ satisfy
\eqn\stand{[F_i, F_j]=i f_{ijk} F_k,} 
with
$f_{123}=1,$
$f_{147}=1/2,$
$f_{156}=-1/2, $
$f_{246}=1/2, $
$f_{257}=1/2, $
$f_{345}=1/2, $
$f_{367}=-1/2, $
$f_{458}=\sqrt{3}/2, $
$f_{678}=\sqrt{3}/2.$

\appendix{B}{Conformal Killing vectors for the $SU(2,1)/U(2)$ 
boundary}
The conformal Killing vectors of the boundary $\xi^k$
should satisfy the following equation
\eqn\ckvb{ {\cal L}_{\xi} \sigma_3=\hat{f}
(\theta,\varphi,\psi) \sigma_3,}
where ${\cal L}$ is the Lie derivative along the
vector field $\xi$.
In components, \ckvb\ takes the form
\eqn\compon{
\partial_{\theta}\xi^{\psi}+\cos{\theta}\partial_{\theta}\xi^{\varphi}=0,}
\eqn\kileqt{-\sin{\theta} \xi^{\theta}+
\partial_{\varphi}\xi^{\psi}+\cos{\theta}\partial_{\varphi}\xi^{\varphi}=
\hat{f} \cos{\theta},}
\eqn\kileqtr{
\partial_{\psi}\xi^{\psi}+\cos{\theta}\partial_{\psi}\xi^{\varphi}=\hat{f}.}
The solution of the above equations with 
$\hat{f} =f e^{\alpha \varphi+\beta \psi}\neq 0$
is given by the following set of vectors,
parametrized by two numbers $\alpha, \beta$ 
and a function $f(\theta),$
\eqn\solckv{
L^{\alpha \beta}_{f}={e^{\alpha \varphi+\beta \psi} \over \beta \sin{\theta}}
[ f(\theta)(\alpha -\beta \cos{\theta}) \partial_{\theta}
-f_{\theta}(\theta)\partial_{\varphi}+
(f(\theta) \sin{\theta}+f_{\theta}(\theta)\cos{\theta} )
\partial_{\psi}].}
For $\hat{f}=0$ we get
\eqn\adisom{
H^{a}_{g}=e^{a \varphi}[
-{ a \over \sin{\theta}}\{ \int d\theta g(\theta) \sin{\theta}+ C \}
\partial_{\theta} + g(\theta)\partial_{\varphi}-
\{\int d\theta \cos{\theta}g_{\theta}(\theta) +C\}\partial_{\psi}.
]}

Here  $\alpha$, $\beta$, $a$ and $C$ are arbitrary constants and 
$f(\theta),$ $g(\theta)$ are arbitrary functions
of $\theta$ such that the corresponding Killing vectors
are nonsingular.
These generators enlarge the $SU(2,1)$ algebra of conformal Killing vectors. 

\listrefs

\end